\documentclass[twocolumn,10pt]{article}

\usepackage[a4paper,margin=0.5in]{geometry}
\usepackage{authblk}
\usepackage{graphicx}
\usepackage{amsmath,amssymb}
\usepackage{cite}
\usepackage{hyperref}
\usepackage{abstract}
\usepackage{caption}
\usepackage{titlesec}

\titlespacing{\section}{0pt}{12pt}{6pt}

\title{\bfseries Imaging domain boundaries of rubrene thin crystallites by photoemission electron microscopy}
\author[1,2]{Moha Naeimi}
\author[1,2]{Katharina Engster}
\author[1,2]{Waqas Pervez}
\author[1,2]{Ingo Barke}
\author[1,2]{Sylvia Speller}
\affil[1]{\small Institute of Physics, University of Rostock, Albert Einstein Straße 23, Rostock, Germany}
\affil[2]{\small Department Life, Light \& Matter, University of Rostock, Albert Einstein Straße 25, Rostock, Germany}
\date{}

\begin{document}
\twocolumn[
\maketitle
\vspace{-2em}
\begin{onecolabstract}
\noindent
The progress of designing organic semiconductors is extensively dependent on the quality of prepared organic molecular assemblies, since the charge transport mechanism is strongly efficient in highly ordered crystals compared to amorphous domains. Here we present a comprehensive photoemission electron microscopy (PEEM) and time-of-flight (TOF) spectroscopic study of rubrene ($\mathrm{C}_{48}\mathrm{H}_{24}$) thin crystals focusing on recently developed orthorhombic crystalline morphologies applied in organic electronic devices. Using femtosecond pulsed lasers with photon energies between 3–6 eV, we explore the interplay between photoemission processes, crystal morphology, and defect states. In a 2-photon photoemission process (2PPE), the PEEM images reveal dominant emission localized at domain boundaries, indicating strong contributions from trap states. In contrast, in 1PPE nm excitation uniform emission across the crystal surface is observed, highlighting a fundamental difference in photoemission mechanisms.  Furthermore, in the intermediate photon energy range, we identify a nonlinear, non-integer photon order, where mostly the triclinic morphology contributes to the emission, distinguishing it from the orthorhombic phase. These findings provide a new framework for assessing the quality and internal structure of organic semiconductor thin films via wavelength-dependent photoemission imaging and spectroscopy.
\end{onecolabstract}
\vspace{1em}
]
\section{Introduction}

Organic semiconducting molecules have attracted a lot of attention in the recent decades due to their promising charge transport properties. Among all the organic molecules, rubrene ($\mathrm{C}_{48}\mathrm{H}_{24}$) offers one of the highest charge carrier mobilities \cite{Zhang2010, Nitta2019}. This charge mobility is associated with a process known as singlet fission \cite{Baronas2022, Bai2014}. Singlet fission is the transformation of an excited singlet state into two triplet state via a non-radiative transition \cite{Finton2019, Zhu2017}. The long lived triplet pair then undergoes a fusion process, either by recombination with each other, limited hopping between molecules through Dexter transfer or long range migration in a random walk scheme \cite{Ryasnyanskiy2011}. 

These processes are subtle in amorphous films \cite{Li2015, Piland2013, Park2007} and a crystalline molecular assembly is beneficial to improve the singlet fission efficiency and migration length of the triplets \cite{Wakikawa2025, Breen2017, Volek2023}. Among all of the crystalline phases of rubrene, the orthorhombic crystalline phase offers not only the highest crystal quality but also the highest efficiency of the above mentioned processes which leads to the highest charge mobility \cite{Choi2019, Hathwar2015, vanderLee2022}.

A well known thin crystalline structure based on the rubrene orthorhombic phase has been prepared and studied extensively \cite{Lee2011, Foggiatto2019, Naeimi2024}. These crystalline structures, usually referred to as orthorhombic platelets \cite{Euvrard2022}, are grown from a center of nucleation towards the edge of the platelet with a somewhat segmented rim \cite{Tan2023, Fielitz2016}. These consist of different domains, typically separated by line defects, and appear with distinct intensity and colour under polarised microscopy due to differences in crystal orientation \cite{Fusella2017}.

Here we use photoemission electron microscopy (PEEM) to study the electron spectra of rubrene orthorhombic domains grown on graphite. We investigate the electron spectra from a 1-PPE (one-photon photoemission) and a 2-PPE process and show that the domain boundaries in such crystalline structures are barriers for exciton migration. This method could be utilised to assess the quality of organic devices that use such crystalline structures.

\section{Experimental section}

Rubrene was grown on highly ordered pyrolytic graphite (HOPG) by high rate heating treatments. The sample preparation was introduced in our previous work \cite{Naeimi2024}. The crystals were investigated by polarisation optical microscopy (POM) (Zeiss Axio lab 5) with a colour camera (Axiocam 305 colour R2) and a bright light LED 10W as light source.

For surface potential investigations we used a force microscope (Park Systems NX20) in sideband Kelvin probe force microscopy (KPFM) mode with conductive tips made of chromium platinum (Cr-Pt) exhibiting a cantilever spring constant of 3 N/m and a free eigenfrequency of 75 kHz.

The time-of-flight photoemission electron spectroscopy was conducted in a PEEM (Focus IS-PEEM) at a base pressure of $10^{-10}$ mbar. We used different light sources for photoemission: (1) The 2nd, 3rd and 4th harmonics of a tunable Ti:Sa femtosecond (fs) laser (Mira 900F) yielding photons with energies of 3.1 eV, 4.2 eV, and 6.2 eV, respectively. The repetition rate was 1 Mhz and pulse duration about 150 ps. The light angle of incidence was 23 degrees, ensuring a good alignment of the light polarisation and transition dipole moment of rubrene crystals, which is perpendicular to the substrate.

Visualization and analysis were done using Gwyddion \cite{Neas2011} and Igor Pro (Wavemetrics).

\section{Results and discussion}

\begin{figure}
    \includegraphics[width=1\linewidth]{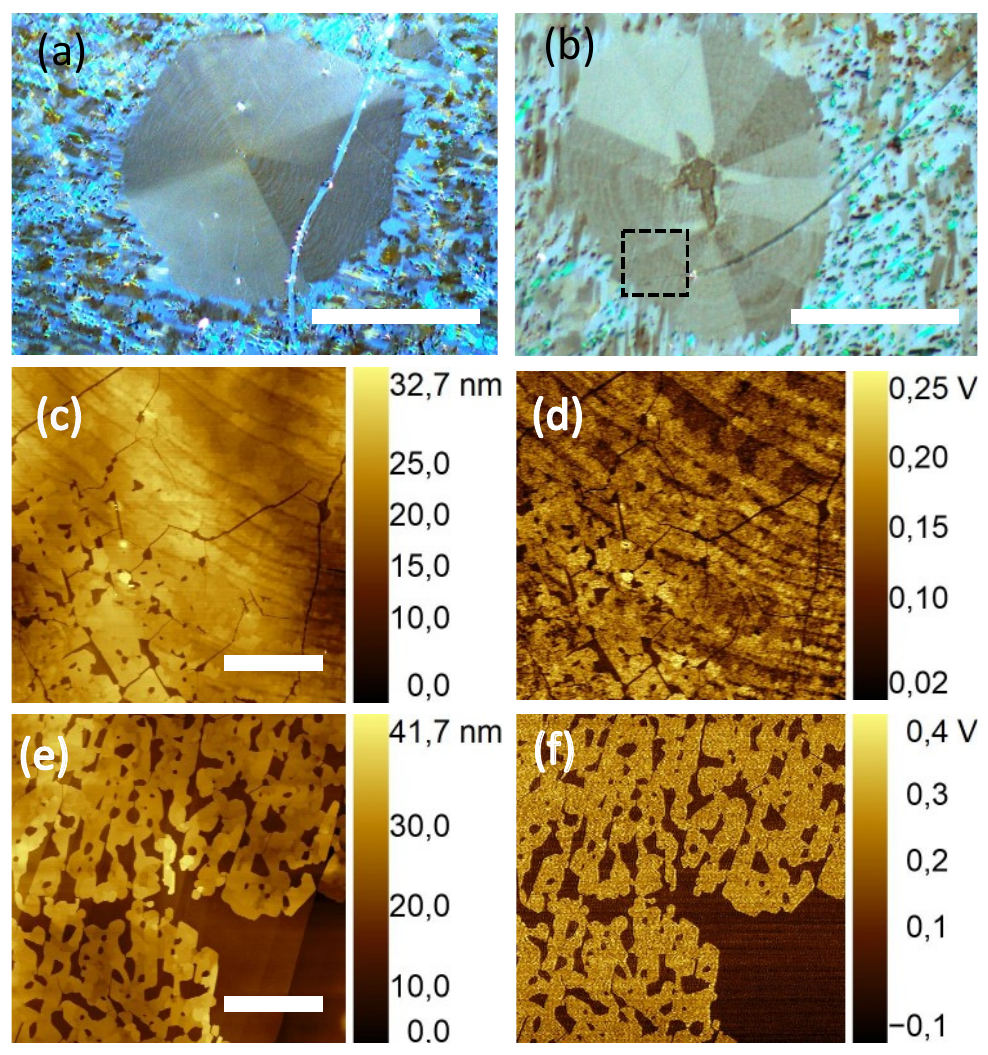}
    \caption{Rubrene polycrystalline structures prepared on HOPG. The orthorhombic platelets with approximately circular shape consist of different domains as visible in polarised optical microscope. They are embedded in a colony of small triclinic crystals (e.g. as seen close to the borders of Fig.XXa), forming triclinic spherulites (not shown, see~\cite{Naeimi2024}). (a and b) Magnified POM images of two rubrene orthorhombic platelet prepared on HOPG. The platelet consists of domains that appear with different colour and brightness in POM. (scale: 100 $\mathrm{\mu m}$). (c and d) AFM and KPFM map of the area marked with a dashed frame in b (scale: 10 $\mathrm{\mu m}$). (e and f) AFM and KPFM map of an area with triclinic crystals. (Scale: 10 $\mathrm{\mu m}$)}
    \label{fig:pol_kpfm}
\end{figure}

Figure \ref{fig:pol_kpfm}a and \ref{fig:pol_kpfm}b show POM images of two different rubrene orthorhombic platelets. The laterally extended thin crystals exhibit a smooth surface with a growth habit corroborating nucleation and growth from the center to the edges of the platelet. These platelets are surrounded by colonies of small crystals forming triclinic spherulites \cite{Naeimi2024, Euvrard2022} (not shown here). 

The contrast observed in polarised optical microscopy (POM) is sometimes linked to line defects formed during sample preparation (Figure \ref{fig:pol_kpfm}b), but it can also appear within a single domain in the absence of such defects (Figure \ref{fig:pol_kpfm}a). It had been shown that these crystals grow with the c-axis of the orthorhombic unit cell perpendicular to the surface \cite{Euvrard2022, Naeimi2024}. This makes the ab-plane of the unit cell lying parallel to the crystal surface and transition dipole moment of S0-S1 normal to the surface \cite{Irkhin2012}.

Figure \ref{fig:pol_kpfm}c and \ref{fig:pol_kpfm}d are atomic AFM and KPFM maps of the dashed framed in Figure \ref{fig:pol_kpfm}b. The line defects between domains are visible in the AFM topography. We measured the KPFM potential difference of rubrene crystals with varying thicknesses and morphologies, and found that the potential does not depend on thickness but varies with morphology, specifically with different molecule density as well as unit cell orientations \cite{Wu2016}.

The surface potential difference variation within the orthorhombic platelets is 100 to 150 meV. The lower local potentials could point to incomplete equilibration of last layer and lower molecule density which is corroborated by topography measurements (Figure \ref{fig:pol_kpfm}c and \ref{fig:pol_kpfm}d). Such KPFM contrast points towards a non-uniform molecular density throughout the crystal surface. The triclinic structures in the surrounding exhibit relatively homogeneous surface potential differences of $\sim$ 400 mV, compared to $\sim$ 200 mV on the HOPG substrate (Figure \ref{fig:pol_kpfm}e and \ref{fig:pol_kpfm}f).

In general, with increasing molecular density, the surface potential drops and the work function accordingly rises; the reason is that higher species density coming along with stronger vertical surface dipole. In ambient experiments, HOPG as well as thin films of aromatic molecules can adsorb airborne molecules with time, which will change the surface potential (\cite{MartinezMartin2013, Terada2021}). Additionally, ambient light can give rise to charge carrier generation, giving rise to surface potential modification.

\begin{figure}
    \includegraphics[width=1\linewidth]{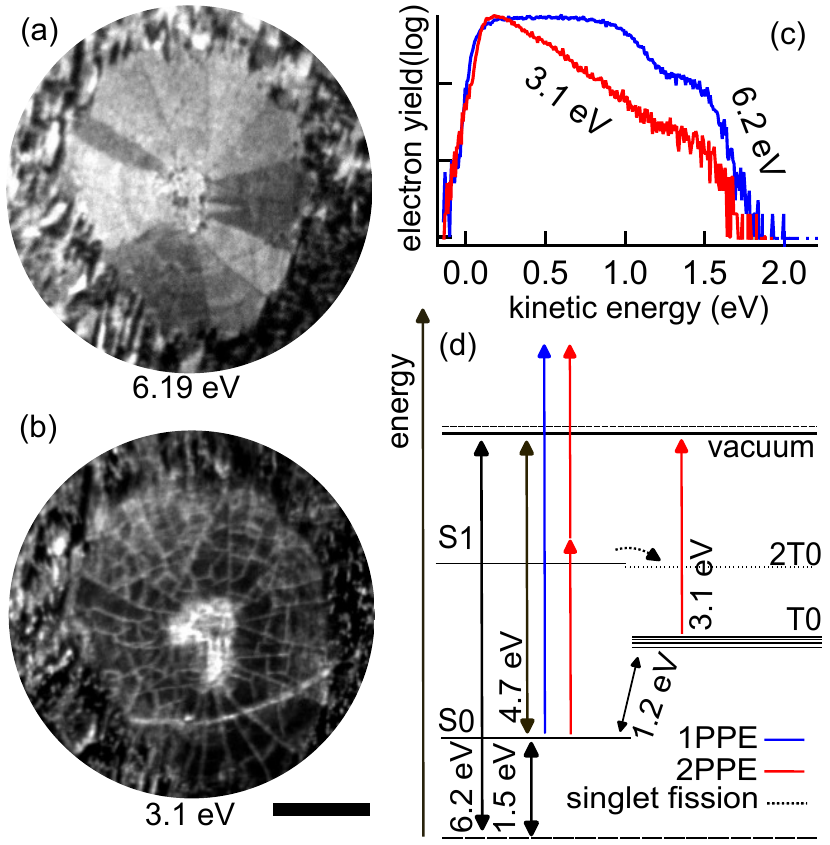}
    \centering
    \caption{1PPE and 2PPE image and electron kinetic energy spectra of rubrene polycrystalline assemblies reveal a fundamental difference in internal photoemission properties. (a and b) PEEM images of the crystal shown in \ref{fig:pol_kpfm}b excited with 3.1 eV and 6.2 eV, respectively. (c) Electron spectra of rubrene orthorhombic platelet excited with different photon energies shown in a and b. (d) Energy diagram of rubrene orthorhombic platelet showing the 1PPE and 2PPE photoemission pathways. (scale: 50 $\mathrm{\mu m}$)}
    \label{fig:PEEM1}
\end{figure}

Figure \ref{fig:PEEM1}a and \ref{fig:PEEM1}b show the PEEM images of the crystal shown in \ref{fig:pol_kpfm}b, excited with different photon energies (6.2 eV and 3.1 eV respectively). Once the crystal is excited with 3.1 eV, the electron yields in the domain boundaries and line defects are higher than in regions with smooth and uniform morphologies, resulting in  a pattern reminiscent to a spider web in Figure \ref{fig:PEEM1}.b.

On the other hand, the 6.2 eV excitation results in a contrast in different domains and local regions of crystals. Unlike the 3.1 eV excitation, domain boundaries and line defects are not appearing bright with high electron yield. 

We attributive the pronounced electron emission from domain boundaries (see \ref{fig:PEEM1}b) to threshold emission from triplet exciton states accumulating at these locations, either via migration or by direct excitation. These states act like trap states which are populated by a first photon. Such processes are associated with a singlet fission upon absorption of the first photon~\cite{Williams1990, Akselrod2014}, followed by subsequent triplet exciton transfer.

Direct electron ejection from occupied states \cite{Okaue2021} as opposed to singlet fission, exciton migration and subsequent emission is much less likely for the 2PPE process compared to 1PPE. The latter results in a contrast governed by different local crystal orientations.

Since the c-axis of the orthorhombic unit cell is oriented along the dipole moment of HOMO-LUMO transition and in our crystals is almost perpendicular to the surface, a local rotation of this axis with respect to the surface normal yields a different angle between excitation polarisation and the dipole moment leading to different electron yields. 

The electron spectra reflect the distinct nature of the involved photoemission processes. Figure \ref{fig:PEEM1}c shows the electron spectra for two excitation sources with photon energies of 3.1 eV and 6.2 eV. The total width of the spectra is same for both 1PPE and 2PPE as expected, but similar kinetic energies point to different states, as pointed out in the following.

In a one-photon photoemission (1PPE) process, the electrons exhibiting the highest kinetic energies must originate from the highest occupied molecular orbital (HOMO), as higher-order excitations are absent. Therefore, the rightmost peak in the 1PPE spectrum can be attributed to the S0 state.

The total spectral width, which is approximately 2 eV thus corresponds to the energy difference between the state with the highest binding energy and the S0 state. The energy difference between the vacuum level and the state with the highest binding energy is equal to the photon energy used for excitation (6.2 eV). Based on this, the position of the vacuum level relative to the HOMO (S0) state is determined to be 4.2 eV above it, see Figure \ref{fig:PEEM1}.

In the two-photon photoemission (2PPE) process, absorption of the first 3.1 eV photon initiates singlet fission, resulting in the formation of two triplet excitons, each with an energy approximately half that of the initial singlet (S0-S1) energy. Consequently, the energy difference between the vacuum level and the triplet state can be as large as 3 eV. Upon absorption of the second 3.1 eV photon in the 2PPE process, electrons emitted from the triplet state are expected to exhibit the lowest kinetic energies.

Accordingly, the leftmost peak in the 2PPE spectrum can be assigned to photoemission from the triplet state. The prominence of this peak points to a high density of trap states associated with domain boundaries and line defects within the crystal, as visible also in Figure \ref{fig:PEEM1}b.

Consequently, the rightmost peak in the 2PPE spectra is originating from an true 2-photon absorption which results in an electron emission having kinetic energies at $\sim 1.5 eV$. The assigned photoemission processes are summarised in Figure \ref{fig:PEEM1}.d. It is important to note that the energy resolution of our setup is approximately 50~meV. We estimate the work function of the orthorhombic rubrene crystal phase to be 4.2~eV. In addition, the time-to-energy conversion introduces an uncertainty of the order of 100~meV, which is comparable to the observed work function variations between local domains ($\sim$ 200~meV).

Moreover, the spatial resolution of the of about 50 nm under the employed experimental conditions limits our ability to resolve photoemission signals from individual domains. As a result, the accuracy of energy assignments is further limited to $\sim$ 200~meV.

\begin{figure} [h]
    \includegraphics[width=1\linewidth]{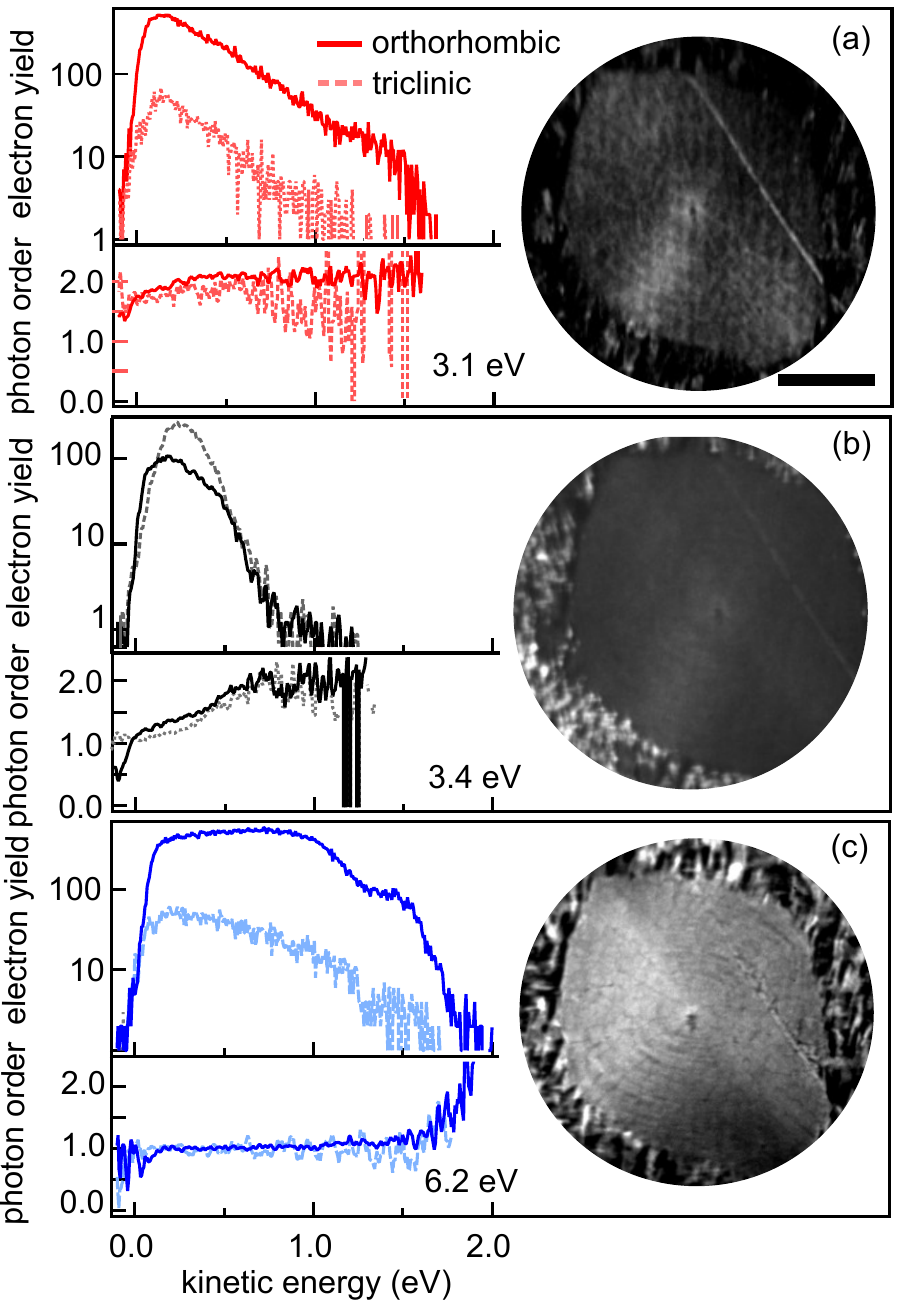}
    \centering
    \caption{(a) Electron spectra (top left) and spectrally resolved photon order (bottom left), and corresponding PEEM image (right) of the rubrene polymorph shown in \ref{fig:pol_kpfm}.a,  excited with photon energies equal to 3.1 eV and resolved for orthorhombic and triclinic crystal phases. (b and c) same as a for excitation photon energies equal to 3.4 eV and 6.2 eV. (scale: 50 $\mathrm{\mu m}$) }
    \label{fig:imspec}
\end{figure}

Nevertheless, PEEM utilising a two-photon photoemission (2PPE) process proves to be a powerful technique for assessing the quality of such crystals, particularly by revealing domain boundaries and defect-induced trap states that are not detectable with conventional polarised optical microscopy. As shown in Figures \ref{fig:pol_kpfm}a and \ref{fig:pol_kpfm}b, the line defects and domain boundaries are not visible; instead, the observed colour variations correspond to different crystal orientations. In contrast, Figures \ref{fig:PEEM1}b and \ref{fig:imspec}a—PEEM 2PPE images of the samples shown in Figures \ref{fig:pol_kpfm}b and \ref{fig:pol_kpfm}a, respectively—clearly reveal pronounced defect lines and domain boundaries, providing direct insight into the crystal quality.

As mentioned above, the emission from occupied states such as triplet or triplet-derived trap-like states implies a second-order process which is investigated in the following by analysing the photon order for different photon energies. Figure \ref{fig:imspec} shows the electron spectra of the crystal shown in \ref{fig:pol_kpfm}a, resolved different crystal phases, excited with different photon energies. For electrons emitted with different kinetic energies, the photon order is also measured.

The photon order remains consistently equal to two across the entire spectral range when the sample is excited with 3.1 eV photons (Figure \ref{fig:imspec}a), except for at lowest kinetic energies. The latter correspond to threshold emission, where the initial states are the above mentioned defect states at the domain boundaries. The reduction of the photon order points to partial saturation of these states and supports the energetic scheme in \ref{fig:PEEM1}d.

For 6.2 eV excitation (Figure \ref{fig:imspec}c), the photon order is constantly equal to 1. Only for electrons with highest kinetic energies, the photon order apparently approaches two. If such processes are involved, they indeed are expected at the highest kinetic energies. Alternatively, such an apparent second-order process could be due space-charge effects which tend to be most pronounced at the highest energies as well. Both possibilities can arise when the sample is excited with relatively intense light.

In an intermediate regime—where the excitation photon energies are above the 2PPE threshold but still below the work function of rubrene—a mixed photon order is observed. As shown in Figures \ref{fig:imspec}b , the photon order evolves from one to two with increasing kinetic energy, when the excitation photon energy is 3.4 eV. For such intermediate photon energies, the extracted photon order deviates from integer values, falling between one and two suggests the involvement of multi-step excitonic dynamics rather than a pure multiphoton absorption process. At photon energies below the work function, an intermediate state may be populated,  which can give rise to fractional photon orders if it is partially saturated.

An interesting observation is that the orthorhombic platelet is contributing to the photoemission only if there is an integer photon order and for a non-integer order, the crystal shows up dark and the signal is mostly coming from the needle-shape triclinic small crystals around the large domain (Figure \ref{fig:imspec}.b).

Another contributing aspect for a non-integer photon order could be the non-uniformity of the work function across the orthorhombic crystal surface or among individual triclinic microcrystals. Such variations in local work function can lead to differences in the local photoemission order. Since the measured signal averages photoemission from the entire surface, encompassing multiple polymorphs, this results in an apparent non-integer photon order. This interpretation is in agreement with the observed increase in photon order with kinetic energy (see Figure \ref{fig:imspec}b and \ref{fig:imspec}c).

\section{Conclusion}
The photoemission signature and electron spectra of rubrene orthorhombic platelets prepared on HOPG were investigated. We showed that wavelength-dependent photoemission electron microscopy, combined with time-of-flight spectroscopy, provides a powerful approach to Figure out the internal electronic mechanisms and morphological quality of rubrene thin crystals. Our findings reveal that two-photon photoemission processes are highly sensitive to trap states and domain boundaries, while one-photon photoemission enables uniform surface imaging. The role of triplet excitons in two-photon emission processes was identified, offering new insight into charge transport dynamics in organic semiconductors. 
Additionally, the observation of a non-integer photoemission order for certain excitation photon energies, highlighting triclinic morphologies over orthorhombic, suggests that spectrally and spatially resolved photoemission could be applied for distinguishing crystalline phases. Surface potential maps of these crystals obtained by KPFM suggest that, such non-integer orders at certain photon energies arise due to spatial variations in the work function. These results establish a framework for quality assessment of organic electronic materials improving device fabrication and characterization.

\section*{Acknowledgment}

Funding by the Deutsche Forschungsgemeinschaft
(DFG, German Research Foundation) within project
number INST 264/110-1 FUGG, SFB 1477 ”Light-
Matter Interactions at Interfaces” and project number
441234705, SFB 1270 ”Electrically Active Implants” is
acknowledged. We thank Sasankan Vinod Kumar and Gyanee Sita Babooram for their help with KPFM measurements.

\bibliographystyle{unsrt}
\bibliography{ref}

\end{document}